\documentstyle{mn}
\input{epsf}

\title[Dark matter distribution in the Coma cluster]
{Dark matter distribution in the Coma cluster from galaxy kinematics:
breaking the mass-anisotropy degeneracy}
\author[Ewa L. {\L}okas and Gary A. Mamon]{Ewa
    L. {\L}okas$^{1}$\thanks{E-mail: lokas@camk.edu.pl}
    and Gary A.
    Mamon$^{2,3}$\thanks{E-mail: gam@iap.fr}\\ $^1$Copernicus Astronomical
    Center, Bartycka 18,
    00-716 Warsaw, Poland\\ $^2$Institut d'Astrophysique de Paris
   (CNRS UMR 7095), 98 bis Bd Arago, F-75014 Paris, France \\ $^3$GEPI
    (CNRS UMR 8111), Observatoire de Paris,
    F-92195 Meudon, France}

\def\farcm{\hbox{$.\mkern-4mu^\prime$}}
\begin{document}

\maketitle

\begin{abstract}
We study velocity moments of elliptical galaxies in the Coma cluster using Jeans
equations. The dark matter distribution in the cluster is modelled by a generalised
formula based upon the results of cosmological $N$-body simulations. Its
inner slope (cuspy or flat), concentration, and mass within the
virial radius are kept as free parameters, as well as the velocity
anisotropy, assumed independent of position. We show that the study of line-of-sight
velocity dispersion alone does not allow to constrain the parameters. By a joint analysis
of the observed profiles of velocity dispersion and kurtosis we are able to break the
degeneracy between the mass distribution and velocity anisotropy. We determine the dark matter
distribution at radial distances larger than 3\% of the virial radius and we find that the galaxy
orbits are close to isotropic. Due to limited resolution, different
inner slopes are found to be consistent with the data and we observe a strong degeneracy
between the inner slope $\alpha$ and concentration $c$: the best-fitting profiles
have the two parameters related with $c=19 - 9.6 \,\alpha$. Our best-fitting NFW profile
has concentration $c=9$, which is 50\% higher than standard values
found in cosmological simulations for objects of similar mass. The total mass within the
virial radius of $2.9\,h_{70}^{-1}$ Mpc is $1.4 \times 10^{15}
h_{70}^{-1}\, M_{\sun}$ (with 30\% accuracy), 85\% of which is dark.
At this distance from the cluster centre, the mass-to-light ratio in the blue
band is $351\,h_{70}$ solar units. The total mass within the virial radius leads to estimates
of the density parameter of the Universe, assuming that clusters trace the
mass-to-light ratio and baryonic fraction of the Universe, with
$\Omega_0=0.29 \pm 0.1$.

\end{abstract}


\section{Introduction}

The Coma cluster of galaxies (Abell 1656) is one of the most extensively studied
in our neighbourhood (see e.g. Biviano 1998 and references therein). Starting with
the seminal paper of Kent \& Gunn (1982) significant effort went into dynamical
modelling of the cluster. In the early studies based on about 300 galaxy velocities only
velocity dispersion was modelled and it was most often assumed that the mass follows
light and that the galaxies are on isotropic orbits. Merritt (1987) showed that if
a larger variety of models is allowed there is a strong degeneracy between the dark
matter distribution and velocity anisotropy and many models can be shown to be consistent
with the data. Without any prior knowledge on the mass distribution even considering
higher velocity moments would probably not be of much help.

Recently, due to theoretical progress mainly by the means of $N$-body simulations,
our knowledge on possible dark matter distributions within gravitationally bound
objects has improved significantly. There seems to be general agreement at least
as to the behaviour of dark matter density profiles at large radial distances ($\varrho
\propto r^{-3}$). Whether the inner dark matter density profile is $\varrho
\propto r^{-1}$ (as in the
so-called universal profile advocated by Navarro, Frenk \& White 1997,
hereafter NFW)
or $\varrho \propto r^{-3/2}$ (as preferred by Moore et al. 1998; see also
Fukushige \& Makino 1997) or is flat (as suggested by
the observed rotation curves of dwarf and low surface brightness galaxies,
e.g. McGaugh \& de Blok 1998) is still a matter of debate (a
recent analysis by Jimenez, Verde \& Oh (2003) of high resolution rotation curves
of spiral galaxies shows
that 2/3 of the sample can be accounted by NFW profiles, but 2/3 also with a
flat core). We
therefore consider a generalised profile with different inner slopes and also
allow for different dark matter concentrations. We constrain this variety
of dark matter density profiles by modelling velocity moments of galaxies. In addition to
constraints from the line-of-sight velocity dispersion profile, we
incorporate constraints from
the fourth velocity moment, the kurtosis.

The NFW profile has been found consistent with the total mass distribution inferred
from the galaxy data combined from many clusters in the {\sf CNOC1} (van der Marel
et al. 2000) and {\sf ENACS} (Biviano et al. 2003) surveys. Although van der Marel
et al. (2000) considered higher velocity moments they did not apply them
rigorously to further constrain the mass distribution. Studies based upon
X-rays, assuming that the hot X-ray emitting gas is in hydrostatic
equilibrium in a spherical potential, usually lead to NFW-like cuspy centres
(McLaughlin 1999; Tamura et al. 2000; Sato et al. 2000). The studies based on gravitational
lensing
focus on the inner shape of the density profile. The slopes agree with
the NFW prediction in some studies (e.g. Broadhurst et al. 2000) while one
team finds a
preference for a flat core (Tyson, Kochanski \& dell'Antonio 1998, see also Williams,
Navarro \& Bartelmann 1999).
Note that the Coma cluster is too close for its mass profile to be probed
through gravitational
lensing.

The amount of galaxy velocity as well as brightness and morphological type measurements
for the members of Coma has increased over the last two decades making it possible
to analyse separately the samples of elliptical and spiral galaxies. While ellipticals
appear to be in dynamical equilibrium justifying the application of Jeans formalism
to study their velocity moments, most of the spirals are probably infalling onto the
cluster. The modelling of the infall of spirals will be presented in the follow-up study.

The paper is organised as follows. In Section~2 we describe our data. In Section~3
we present our assumptions concerning the matter content of the cluster, i.e.
the distributions of galaxies, gas and dark matter. Section~4 outlines
our formalism for modelling the velocity moments of elliptical galaxies
based on Jeans equations. Section~5 presents its application to constrain the
parameters of our model, in particular the distribution of dark matter in the cluster.
The discussion follows in Section~6.

\section{The data}

\begin{figure}
\begin{center}
    \leavevmode
    \epsfxsize=8cm
    \epsfbox[45 40 330 790]{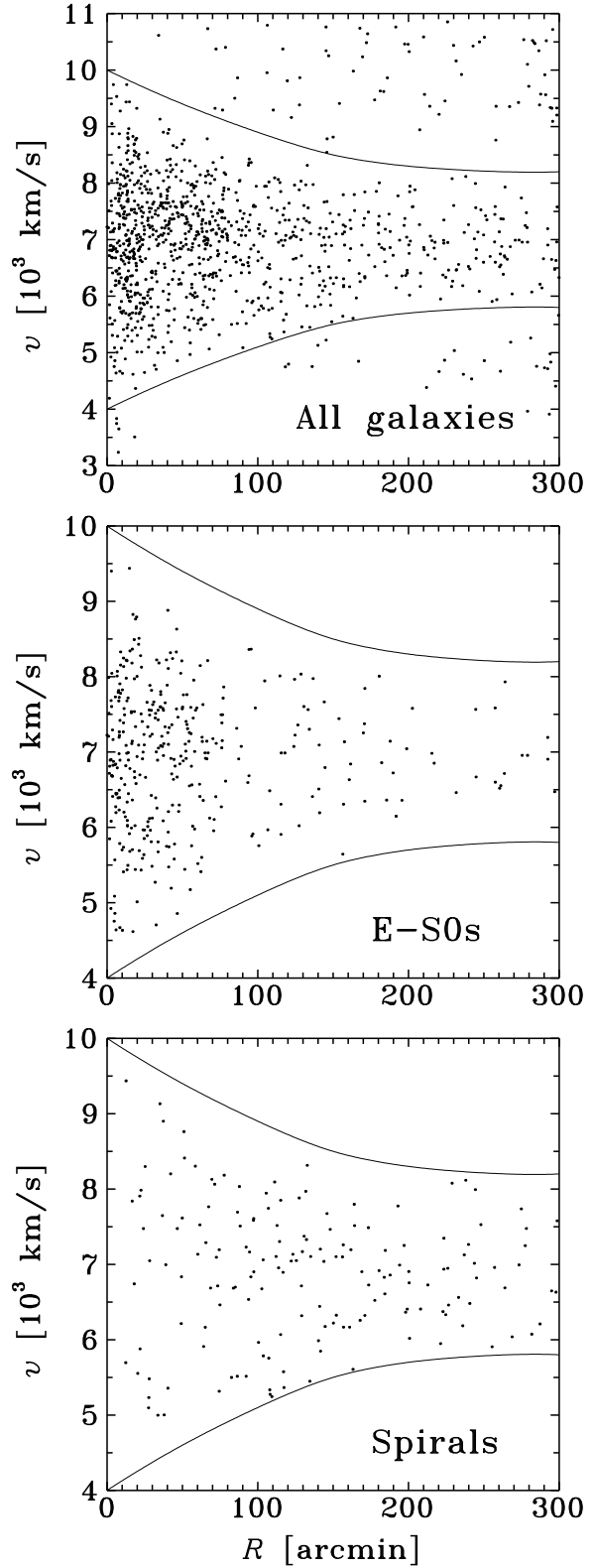}
\end{center}
\caption{Upper panel: 1068 galaxies selected from the {\sf NED} database
within $300'$
from NGC 4874 with heliocentric velocities between $3000$ and $11000$ km/s. Middle panel:
355 E-S0 galaxies, members of Coma. Lower panel: 163 spiral galaxies, members
of Coma. The curves indicate envelopes of the cluster.}
\label{gal}
\end{figure}

We have searched the {\sf NED} database for galaxies within $300'$ of
RA=12h59m35.7s, Dec=$+27^\circ57'33''$ (J2000) i.e. the position of the
elliptical galaxy NGC 4874, well established as the centre of the Coma
cluster (Kent \& Gunn 1982). The galaxies were required to have velocities
between $3000$ and $11000$ km/s (given the velocity dispersions we will find
below, this velocity range extends to $\ga 4\,\sigma$). For the calculation of the velocity
moments and the subsequent study of kinematics, we remove from the list
galaxy pairs and known
members of pairs (as given by {\sf NED}), as we wish to probe the global cluster potential but not
its local enhancements.
We then obtain a sample of 1068 galaxies shown in the upper panel
of Figure~\ref{gal} as points in the plane of velocity versus distance
from the centre of the cluster. To determine the membership of galaxies in
the cluster we proceed in a similar fashion as Kent \& Gunn (1982).
As can be seen from Figure~\ref{gal}, the members
of the cluster are well separated from the foreground and background galaxies in
velocity space. We have therefore selected the probable members of Coma as
lying within the two curves shown in Figure~\ref{gal},
symmetric with
respect to $v=7000$ km/s, the value close to the mean velocity of the
cluster. This procedure leaves us with 967 galaxies. For the determination
of the luminosity distribution, we keep the members of galaxy pairs and
remove a few galaxies for which no magnitude estimate is available (they
may contribute to the calculation of velocity moments however). We then
proceed with the membership determination as before.

It is generally believed that only early-type (E-S0) galaxies can be considered
in dynamical equilibrium within a cluster in opposition to spirals which
are believed to be infalling (e.g. Tully \& Shaya 1984; Huchra 1985). We therefore construct
separate samples of E-S0 and spiral galaxies.
The morphological type of the galaxies has been determined by consulting
the {\sf NED}, {\sf SIMBAD} and {\sf LEDA} databases. Among the 967 galaxies belonging to Coma
and selected for the analysis of the velocity moments, we
find $355$ E-S0s, $163$ spirals and $449$ other galaxies for which
the morphological classification is unknown or uncertain. As already
noticed by Kent \& Gunn (1982), the cluster shows a clear morphological
segregation. The distributions of E-S0s and
spirals in the $R-v$ plane are shown in the middle and lower panel of
Figure~\ref{gal}. Comparison of the two plots reveals that
while E-S0s cluster and dominate in the central
parts of Coma and are not numerous at radial distances larger than $80'$,
spirals are more uniformly distributed and underrepresented in the
central region. Similar subsamples are constructed for the analysis of
the luminosity distribution.
Note that iterating on the mean velocity and establishing new symmetric
envelopes for
the cluster in the observed phase space has virtually no effect on the cluster
membership. We note a possible group of galaxies at $R < 10'$ and $v < 4000$
km/s that might have been missed, but contributes little to
the central internal kinematics.

\section{The matter content of the cluster}

\subsection{The mass in stars}

The mass contributed by the stars in galaxies is estimated as follows.
Ideally, one would like to determine the surface
luminosity distribution separately for different morphological types of galaxies, since
transforming luminosity into mass requires mass-to-light
ratios, which are known to vary with morphological type. However, since we do
not know the morphological type for roughly half
the galaxies in our sample, we would have to assume for these galaxies some mean value of
the mass-to-light ratio.
Combining then the fits for the three classes of
galaxies: E-S0s, spirals and those of unknown type would produce large
uncertainties, since the luminosity distribution for spirals turns out
to be quite noisy.
%Moreover, the small number of spirals would lead to an uncertain luminosity
%distribution for late-type galaxies.
Therefore, in determining the total stellar mass distribution,
we fit the luminosity distribution for all galaxies
and then translate it to the mass distribution using a mean mass-to-light
ratio.
%The drawback is that morphological segregation will lead to a
%flatter and more extended distribution for the spirals, which is not taken
%into account in our analysis.

Magnitudes are transformed into luminosities assuming that all
galaxies are at the same distance associated with the mean velocity of the
Coma cluster. Adopting an heliocentric velocity of Coma of 6925 km/s
(Struble \& Rood 1999), and correcting for the velocity of the Sun with
respect to the Local Group and for the Local Group infall onto Virgo, we
obtain for Coma a Hubble flow velocity of 7093 km/s, which, for a Hubble constant of
$H_0 = 70 \,\rm km \,s^{-1} \, Mpc^{-1}$
(assumed throughout this paper) gives a distance of 101.3 Mpc and a
distance modulus of 35.03 (neglecting the peculiar velocity of the Coma
cluster).

\begin{figure}
\begin{center}
    \leavevmode
    \epsfxsize=8cm
    \epsfbox[40 40 330 300]{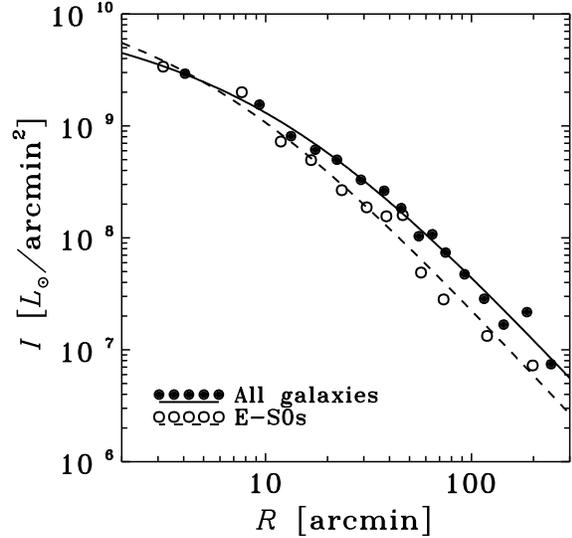}
\end{center}
\caption{The surface luminosity distribution data together with the
best-fitting projected NFW profiles for all 985 galaxies (filled symbols and solid
line) and the 366 known E-S0s among them (open symbols and dashed line).}
\label{surlum1}
\end{figure}

As mentioned in the previous section, our sample of galaxies for the
luminosity analysis is somewhat different from the one used for the
calculation of velocity moments, as we no longer exclude the members of
pairs. Now, we have 985 galaxies that belong
to Coma, among which 366 E-S0s and 167 spirals.
The surface luminosity profile of all galaxies is then determined by placing
60 galaxies per radial bin. The resulting distribution is shown as
filled symbols in Figure~\ref{surlum1}. The open symbols in the
Figure show a similar result for just E-S0 galaxies but this time with 30
galaxies per radial bin. This second distribution will be needed in the
modelling of the velocity moments as the distribution of our tracer
population.

The data shown in Figure~\ref{surlum1} do not indicate any
presence of a core in the surface luminosity distributions, hence we fit them with
projections of cuspy profiles. For both samples, of all galaxies as well as early-type ones,
the distributions have a changing slope, so we fit them with a
projection of the NFW profile (Bartelmann 1996; see also section 2.5 of {\L}okas \& Mamon
2001). The fit is done by $\chi^2$ minimisation taking all points with
equal weights (although we do not know the errors of the magnitude
measurements we expect them to be similar for galaxies in each bin). The
3D luminosity density will then be
\begin{equation}	\label{c1}
	l(r) = \frac{l_\star}{(r/r_{\rm S}) (1+r/r_{\rm S})^2} \ ,
\end{equation}
where the normalisation constant $l_\star$ and the scale radius $r_{\rm
S}$ are the two fitting parameters. For the sample of all galaxies we
obtain $l_\star = 9.05 \times 10^7 L_{\sun}/$arcmin$^3$ and $r_{\rm
S}=14\farcm4$. For the E-S0s we have $l_\star = 3.55 \times 10^8
L_{\sun}/$arcmin$^3$ and $r_{\rm S}=7\farcm05$ so this distribution turns
out to be more concentrated and therefore steeper. It might be interesting
to note that the \emph{number} density distribution of the galaxies in our sample
is less steep in the centre than the \emph{luminosity} density distribution
and can be approximated by a profile with
a core. The steeper distribution of luminosity is probably mainly due to
the presence of bright ellipticals in the centre of the cluster (see
Figure~\ref{gal}).

Integrating the luminosity distribution (\ref{c1}) and multiplying by the
appropriate mass-to-light ratio in the blue band, $\Upsilon$, we obtain the mass
distribution associated with the stars in galaxies
\begin{equation}	\label{c2}
	M_{\rm G}(r) = 4 \pi L_\star \Upsilon r_{\rm S}^3
	\left( \ln \frac{r+r_{\rm S}}{r_{\rm S}} - \frac{r}{r+r_{\rm S}}
    \right)\ .
\end{equation}
We adopt the mass-to-light ratio in blue band for E-S0s of $\Upsilon_{\rm
E}=8 M_{\sun}/L{\sun}$ and for spirals $\Upsilon_{\rm S} = 3
M_{\sun}/L_{\sun}$. For the mass distribution of all galaxies we take the
weighted mean of $\Upsilon$. Since, for galaxies with known
morphological type, the E-S0s and spirals appear in the proportions of
$\nobreak {2.2\,:\,1}$, then assuming that the same proportion holds for the whole
population, the mean mass-to-light ratio amounts to
$\Upsilon_{\rm G} = (2.2 \Upsilon_{\rm E} + \Upsilon_{\rm S})/3.2 =
6.43 M_{\sun}/L_{\sun}$.

\subsection{The gas distribution}

To approximate the contribution of the gas to the mass in the cluster we
make use of the X-ray surface brightness distribution which can be well
approximated by the following formula
\begin{equation}	\label{c3}
    S(r) = S_0  \left[ 1+ \left( \frac{r}{r_{\rm c}} \right)^2 \right]^{-3 b + 1/2}.
\end{equation}
Briel, Henry \& B\"{o}hringer (1992) analysed the {\sf ROSAT} observations of
Coma out to $100'$ in the energy band of 0.5 to 2.4 keV
and obtained the following best-fitting parameters: $S_0 = 4.6
\times 10^{-13}$ erg cm$^{-2}$ s$^{-1}$ arcmin$^{-2}$, $b=0.75$ and
the core radius $r_{\rm c}=10\farcm5$. Assuming that the radiation is
produced by bremsstrahlung this distribution is obtained by projecting the
3D emissivity integrated over the energy band and assuming that the gas is
approximately isothermal. Since the emissivity is proportional to the square
of the number density of electrons in the gas $n$, we obtain
\begin{equation}	\label{c4}
    n(r) = n_0  \left[ 1+ \left(\frac{r}{r_{\rm c}} \right)^2 \right]^{-3 b/2}
\ ,
\end{equation}
where the central electron density (with $h=0.7$) is $n_0 = 3.42 \times 10^{-3}$
cm$^{-3}$ (Briel et al. 1992; Henry \& Henriksen 1986).
Integrating equation (\ref{c4}) and multiplying by the
mass per electron we obtain the mass distribution associated with the gas
\begin{equation}	\label{c5}
    M_{\rm g}(r) = \frac{4}{3} \pi n_0 (m_{\rm e} + \gamma m_{\rm p}) r^3 \  _2 F_1
    \left( \frac{3}{2}, \frac{3 b}{2}; \frac{5}{2}; -\frac{r^2}{r_{\rm c}^2}
    \right) \ ,
\end{equation}
where ${}_2F_1$ is the hypergeometric function and
$\gamma m_{\rm p}$ with $\gamma=1.136$ is the mean mass of the positively
charged ion in the gas per unit charge ($m_{\rm p}$ being the proton mass),
assuming the gas has the primordial
composition with helium abundance of $Y_{\rm He} =0.23-0.24$.

The formula (\ref{c3}) for the surface brightness is the same as the one appearing
in the so-called $\beta$-model (we have replaced $\beta$ by $b$ to avoid confusion
with the anisotropy parameter of the next section), however we do not accept all the
assumptions of the model here concerning e.g. the relation between the gas and galaxy
distributions. The only assumption going into the derivation of the gas mass distribution in
equation (\ref{c5}), besides spherical symmetry, is that the gas is isothermal, which
is justified by recent observations of Coma with {\sf XMM-Newton} by Arnaud
et al. (2001), who find
very little temperature variation at least in the central region of radius $20'$.

\subsection{Dark matter}

We study different possible density distributions of dark matter described
by the following general formula
\begin{equation}    \label{m6}
    \rho(r) = \frac{\rho_{\rm char}}{(r/r_{\rm
    s})^\alpha \,(1+r/r_{\rm s})^{3-\alpha}} \ ,
\end{equation}
where $\rho_{\rm char}$ is a constant characteristic density, while $r_{\rm
s}$ is the scale radius of the dark matter (in general different from that of
the luminous distribution, $r_{\rm S}$). The profiles
differ by the inner slope $r^{-\alpha}$
but have a common outer limiting behaviour of $r^{-3}$.
We will consider $\alpha$ values limited by $0 \le \alpha \le 3/2$, which covers
a wide range of possible inner profiles: from very steep to core-like.
The cuspy profiles of $\alpha>0$ are
motivated by the results of cosmological $N$-body simulations. The profile with $\alpha=1$
corresponds to the so-called universal profile proposed by NFW as a fit to the profiles
of simulated haloes, while the
profile with $\alpha=3/2$ is identical to the one following from higher resolution
simulations of Moore et al. (1998). The core profile with $\alpha=0$
is favoured by some observations of galaxies and clusters and is very similar
(but not identical) to the profile proposed by Burkert (1995).

The scale radius $r_{\rm s}$ introduced in equation (\ref{m6}) marks the distance
from the centre of the object
where the profile has a slope equal to the average of the inner and outer slope:
$r^{-(3+\alpha)/2}$. The other parameter that controls the shape of the profile
is the concentration
\begin{equation}    \label{d1}
    c=\frac{r_v}{r_{\rm s}},
\end{equation}
where $r_v$ is the virial radius, i.e. the distance from the centre
of the halo within which the mean density is $\Delta_{\rm c}$ times the
present critical density, $\rho_{\rm crit,0}$. Although in most cosmological $N$-body simulations
$\Delta_{\rm c}=200$ is assumed and kept constant, the value, following from the
spherical collapse model, depends on the underlying cosmology,
e.g. $\Delta_{\rm c} \approx 178$ is valid for the Einstein-de Sitter model,
while, for the currently
most popular $\Lambda$CDM model with $\Omega_M=0.3$ and
$\Omega_{\Lambda}=0.7$,
 we have $\Delta_{\rm c} \approx 102$
(Eke, Cole \& Frenk 1996; \L okas \& Hoffman 2001).
We will keep $\Delta_{\rm c}=102$ in the following.

The concentration of simulated dark matter haloes has been observed to
depend on the virial mass. Jing \& Suto (2000) tested the relation $c (M_v)$
for the masses of the
order of normal galaxies and clusters in the case of density profiles with
$\alpha=1$ and $\alpha=3/2$ and found concentrations slowly decreasing with mass
(thus confirming the original observation of NFW) and
lower for $\alpha=3/2$ than for $\alpha=1$.
The only study using the value of $\Delta_{\rm c}$ appropriate for a given cosmological model is
the one of Bullock et al. (2001) who found the profiles of presently
forming haloes to be well fitted by
the NFW formula with concentrations depending on mass in the $\Lambda$CDM model
with the above parameters approximately as
\begin{equation}    \label{c6a}
     c(M_v) = 5.95
    \left( \frac{M_v}{10^{15} h^{-1} M_{\sun}} \right)^{-0.122}.
\end{equation}
In the following, we will treat the concentration as a free parameter using this relation
to guide us as to order of magnitude of $c$.

We normalise the density profile (\ref{m6}) so that the mass within $r_v$
is equal to the so-called virial mass
\begin{equation}    \label{d2}
    M_v=\frac{4}{3} \pi r_v^3 \Delta_{\rm c} \rho_{\rm crit,0} \ .
\end{equation}
The characteristic density of equation (\ref{m6}) then becomes
\begin{equation}    \label{d3}
    \rho_{\rm char}=\frac{(3-\alpha) \Delta_{\rm c} \rho_{\rm crit,0} \ c^\alpha}{3
    F_\alpha(c)} \ ,
\end{equation}
where $F_\alpha(c)$ is given by the hypergeometric function
\begin{eqnarray}
    F_\alpha(x) &\!\!\!\!=\!\!\!\!& \ _2 F_1 (3-\alpha, 3-\alpha; 4-\alpha; -x)
\label{d4}
\\
&\!\!\!\!=\!\!\!\!&
\left\{\!\!
\begin{array}{ll}
\displaystyle
{3\over x^3}\,\left [\ln(1\!+\!x)-{x\over 2}\, {(2 + 3\,x) \over (1\!+\!x)^2}\right] &
\!\!(\alpha=0) \\
\displaystyle
{5\over x^{5/2}}\,\left [\sinh^{-1}\!\sqrt{x} - {\sqrt{x}\over 3}\,{3 + 4\,x\over (1\!+\!x)^{3/2}}\right]
& \!\!(\alpha=1/2)
\\
\displaystyle
{2\over x^2}\,\left[
\ln(1\!+\!x) - {x\over 1\!+\!x}\right] & \!\!(\alpha=1)
\\
\displaystyle
{3\over x^{3/2}}\,\left [\sinh^{-1}\!\sqrt{x} - \sqrt{x\over 1\!+\!x} \right]
& \!\!(\alpha=3/2)
\end{array}
\right.
\nonumber
\end{eqnarray}
The dark mass distribution following from (\ref{m6}), (\ref{d2}) and (\ref{d3}) is
\begin{equation}    \label{d5}
    M_{\rm D}(s) = M_v s^{3-\alpha} \frac{F_\alpha(c s)}{F_\alpha(c)} \ ,
\end{equation}
where we introduced $s=r/r_v$.

\section{Modelling of the velocity moments}

The purpose of this work is to constrain the distribution of dark matter in
the Coma cluster by studying the velocity moments of the population of elliptical
galaxies in the cluster which we believe to be in equilibrium (thus
neglecting radial streaming motions). We defer to a later paper the
kinematical study of spiral galaxies in the context of infall. We will also assume that
the system is spherically symmetric and that there are no net streaming motions
(e.g. no rotation) so that the odd velocity moments vanish.

At second order, the two distinct moments
are $\overline{v_r^2}$ and $\overline{v_\theta^2}=\overline{v_\phi^2}$ which we will
denote hereafter by $\sigma_r^2$ and $\sigma_\theta^2$ respectively.
They can be calculated from the lowest order Jeans equation (e.g. Binney \&
Mamon 1982)
\begin{equation}    \label{m1}
    \frac{\rm d}{{\rm d} r}  (\nu \sigma_r^2) + \frac{2 \beta}{r} \nu
	\sigma_r^2 = - \nu \frac{{\rm d} \Phi}{{\rm d} r} \ ,
\end{equation}
where $\nu$ is the 3D density distribution of the tracer population and $\Phi$ is
the gravitational potential.
We will solve equation (\ref{m1}) assuming the anisotropy parameter
\begin{equation}    \label{d7}
	\beta=1-\frac{\sigma_\theta^2(r)}{\sigma_r^2(r)}
\end{equation}
to be constant with $-\infty < \beta \le 1$. This model covers
all interesting possibilities from radial orbits
($\beta=1$) to isotropy ($\beta=0$) and circular orbits
($\beta \rightarrow - \infty$).

The solution of the lowest order Jeans equation with the boundary condition
$\sigma_r \rightarrow 0$ at $r \rightarrow \infty$ for $\beta$=const is
\begin{equation}	\label{m4b}
	\nu \sigma_r^2 (\beta={\rm const})= r^{-2 \beta}
	\int_r^\infty r^{2 \beta} \nu \frac{{\rm d} \Phi}{{\rm d} r} \ {\rm d}r
\ .
\end{equation}
However, the measurable quantity
is the line-of-sight velocity dispersion obtained from the 3D velocity
dispersion by integrating along the line of sight (Binney \& Mamon 1982)
\begin{equation}    \label{m3}
    \sigma_{\rm los}^2 (R) = \frac{2}{I(R)} \int_{R}^{\infty}
    \left( 1-\beta \frac{R^2}{r^2} \right) \frac{\nu \,
    \sigma_r^2 \,r}{\sqrt{r^2 - R^2}} \,{\rm d} r \ ,
\end{equation}
where $I(R)$ is the surface distribution of the tracer and
$R$ is the projected radius.
Introducing equation (\ref{m4b}) into equation (\ref{m3})
and inverting the order of integration, we obtain
\begin{eqnarray}
    \sigma_{\rm los}^2 (R) &=& \frac{2 G}{I(R)} \int_{R}^{\infty}
    \, {\rm d} x \ \nu (x) M(x) x^{2 \beta - 2}    \label{m4c} \\
    & &  \times \int_{R}^{x}  \,{\rm d} y \left( 1-\beta \frac{R^2}{y^2} \right)
    \frac{y^{-2 \beta +1}}{\sqrt{y^2 - R^2}} \ ,
    \nonumber
\end{eqnarray}
where $M(x)$ is the mass distribution and we used new variables $x$ and $y$
instead of $r$ to avoid confusion. The calculation of $\sigma_{\rm los}$
can then be reduced to one-dimensional numerical integration of a formula
involving special functions for arbitrary $\beta = \rm const$.

It has been established that by studying $\sigma_{\rm los} (R)$ alone
we cannot uniquely determine the properties of a stellar system. In fact,
systems with different densities and velocity anisotropies can produce
identical $\sigma_{\rm los} (R)$ profiles (see e.g. Merrifield \& Kent 1990;
Merritt 1987). It is therefore interesting to
consider higher-order moments of the velocity distribution.
For the fourth-order moments, the three distinct components
$\overline{v_r^4}$, $\overline{v_\theta^4}=\overline{v_\phi^4}$ and
$\overline{v_r^2 v_\theta^2}=\overline{v_r^2 v_\phi^2}$ are related by two
higher order Jeans equations (Merrifield \& Kent 1990).

In order to solve these equations we need additional information about the
distribution function. We will restrict ourselves here to functions which can be
constructed
from the energy-dependent distribution function by multiplying it by a function
of angular momentum $f(E, L) = f_0 (E) L^{-2 \beta}$
with $\beta =\rm const$. The solution of the Jeans equation for the fourth-order moment
(see {\L}okas 2002)
\begin{equation}    \label{d11}
        \frac{\rm d}{{\rm d} r}  (\nu \overline{v_r^4}) + \frac{2 \beta}{r} \nu
	\overline{v_r^4} + 3 \nu \sigma_r^2 \frac{{\rm d} \Phi}{{\rm d} r} =0 \ ,
\end{equation}
is
\begin{equation}	\label{d12}
	\nu \overline{v_r^4} (\beta={\rm const})= 3 r^{-2 \beta}
	\int_r^\infty r^{2 \beta} \nu \sigma_r^2 (r)
	\frac{{\rm d} \Phi}{{\rm d} r} \ {\rm d} r \ .
\end{equation}
By projection, we obtain the line-of-sight fourth moment
\begin{equation}         \label{d13}
    \overline{v_{\rm los}^4} (R) = \frac{2}{I(R)} \int_{R}^{\infty}
    \frac{\nu \,  \overline{v_r^4} \,r}{\sqrt{r^2 - R^2}} \ g(r, R, \beta)
    \,{\rm d} r \ ,
\end{equation}
where
\begin{equation}	\label{d15}
    g(r, R, \beta) = 1 - 2 \beta \frac{R^2}{r^2} + \frac{\beta(1+\beta)}{2}
    \frac{R^4}{r^4} \ .
\end{equation}

Introducing equations (\ref{m4b})
and (\ref{d12}) into (\ref{d13}) and inverting the order of integration, we obtain
{\samepage
\begin{eqnarray}
    \overline{v_{\rm los}^4} (R) &=& \frac{6 G^2}{I(R)}
    \int_{R}^{\infty} \frac{r^{-2 \beta+1}}{\sqrt{r^2 - R^2}}\,g(r, R, \beta)
    \,{\rm d} r  \nonumber \\
    & & \times \int_{r}^{\infty} \frac{\nu(q)\, M(q) }{q^{2 - 2
    \beta}}\,{\rm d} q \int_{r}^{q} \frac{ M(p)}{p^2}\,{\rm d} p
     \label{d16}
\end{eqnarray}
}
and the
calculation of $\overline{v_{\rm los}^4} (R)$ can be reduced to a (rather
long) double integral. A useful way to express the fourth projected moment is to
scale it with $\sigma_{\rm los}^4$ in order to obtain the projected kurtosis
\begin{equation}	\label{d14}
	\kappa_{\rm los} (R) = \frac{\overline{v_{\rm los}^4} (R)}
	{\sigma_{\rm los}^4 (R)} \ ,
\end{equation}
whose value is 3 for a Gaussian distribution.

\begin{figure}
\begin{center}
    \leavevmode
    \epsfxsize=8cm
    \epsfbox[45 40 330 550]{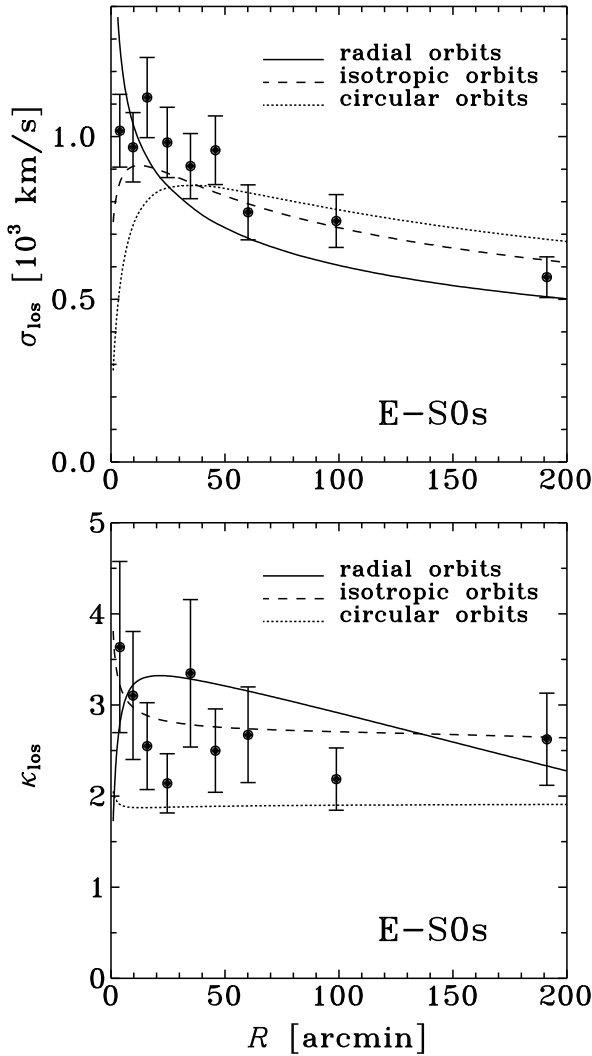}
\end{center}
\caption{The line-of-sight velocity dispersion (upper panel) and the
dimensionless line-of-sight kurtosis parameter (lower panel) of E-S0
galaxies.
The curves represent models with stars, hot gas and
dark matter with NFW distribution of mass $M_v =
10^{15}\,M_\odot$ and concentration $c=6$.
}
\label{vdkurte2}
\end{figure}

\section{Results}

We use our sample of galaxy velocities to calculate the velocity moments of the E-S0
galaxies. The 355 available E-S0
galaxies were divided into bins of 39 objects. Figure~\ref{vdkurte2} shows the
line-of-sight velocity dispersion (upper panel) and kurtosis (lower panel)
together with their sampling errors calculated using estimators based on Monte Carlo
simulations (see Appendix A). We find the sampling distribution of $\sigma_{\rm los}$
to be close to normal. In the case of kurtosis, shown in the lower panel of
Figure~\ref{vdkurte2}
the values are actually calculated and their errors propagated from the quantity
$(\log \kappa_{\rm los})^{1/10}$ which we find to be normally distributed
(see Appendix A). Normal sampling distributions of the estimators of both moments
and very weak correlations between them justify the use of standard fitting procedures
of these quantities based on $\chi^2$ minimisation.

\begin{figure}
\begin{center}
    \leavevmode
    \epsfxsize=8cm
    \epsfbox[45 40 330 790]{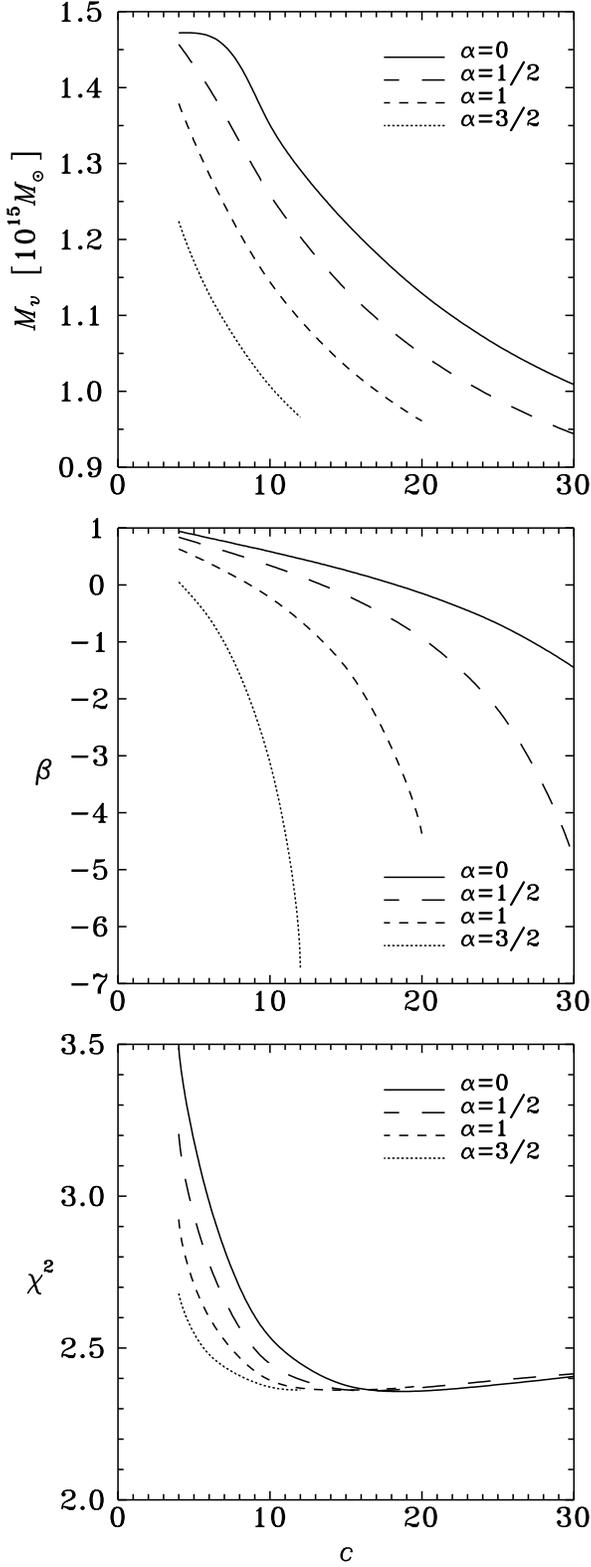}
\end{center}
\caption{Results of fitting line-of-sight velocity dispersion data.
The best fitting parameters $M_v$ and $\beta$ are shown in the two upper
panels as a function of concentration for different $\alpha$. The lower panel
gives the goodness of fit $\chi^2$.}
\label{vdfit}
\end{figure}

Our purpose here is to reproduce the observed velocity moments using models described
by equations (\ref{m3}) and (\ref{d16}), with the mass distribution given by the sum of
the three contributions (\ref{c2}), (\ref{c5}) and (\ref{d5}) discussed in Section~3:
\begin{equation}    \label{c7}
	M(r) = M_{\rm G} + M_{\rm g} + M_{\rm D} \ .
\end{equation}
The density profile $\nu(r)$ of the tracer population of early-type galaxies is given by equation
(\ref{c1}) and the surface brightness $I(R)$ by its projection.
While studying velocity dispersion is useful to constrain the mass, the kurtosis is
mostly sensitive to the velocity anisotropy.

To give a feeling of its behaviour, we show
in Figure~\ref{vdkurte2} the predictions of equation (\ref{d16}) (and the corresponding
ones of equation (\ref{m3})) for the three cases of radial ($\beta=1$), isotropic ($\beta=0$)
and circular ($\beta=-\infty$) orbits assuming dark matter distribution given by an NFW
profile (equation (\ref{m6}) with $\alpha=1$) for the virial mass $M_v=10^{15} M_{\sun}$
and with concentration $c=6$ (as suggested by formula (\ref{c6a}) for the mass of this
order). We see that for radial orbits
the kurtosis profile is convex as opposed to the concave shapes in the case of isotropic
and circular orbits. Since our data seem to prefer a concave shape we can expect
isotropic or tangential orbits to fit the data best.

We begin by fitting the line-of-sight velocity dispersion profile shown in
the upper panel of Figure~\ref{vdkurte2}. We consider different values of
$\alpha=0, 1/2, 1, 3/2$, and, for each of them,
we determine the best-fitting anisotropy parameter $\beta$ and dark virial
mass $M_v$ as a function of concentration $c$. The best-fitting values of
$M_v$ obtained are of the order of
$10^{15} M_{\sun}$, which corresponds to dark halo virial radius $r_v=
88'$ or $2.6$ Mpc. The virialised region is supposed to lie within this radius,
so in the following analysis we discard the two outer radial bins.

Figure~\ref{vdfit} shows the results of fitting just the inner $7$ data points in the
upper panel of Figure~\ref{vdkurte2} for $\alpha=0, 1/2, 1$ and $3/2$.
The best-fitting dark virial mass $M_v$ (upper panel) and the velocity anisotropy parameter
for early-type galaxies $\beta$ (middle panel) are shown as a function of concentration.
The lower panel of Figure~\ref{vdfit} shows the goodness of fit $\chi^2$ (we do not
use $\chi^2/N$, with $N$ the number of degrees of freedom, because it is not obvious
how many parameters can be estimated with this procedure). As can be seen in
Figure~\ref{vdfit}, the biggest $M_v$ obtained is $1.5 \times 10^{15} M_{\sun}$
which gives $r_v=100'$ or $2.9$ Mpc. The data points are now all within this
region and therefore there is no need for further adjustments in the number of data
points analysed. The lowest panel of Figure~\ref{vdfit} proves that neither $c$ nor
$\alpha$ can be constrained from the analysis of velocity dispersion alone:
$\chi^2$ flattens for large $c$ and reaches similar values for all $\alpha$ for
a wide range of $c$.

\begin{figure}
\begin{center}
    \leavevmode
    \epsfxsize=8.5cm
    \epsfbox[130 110 460 590]{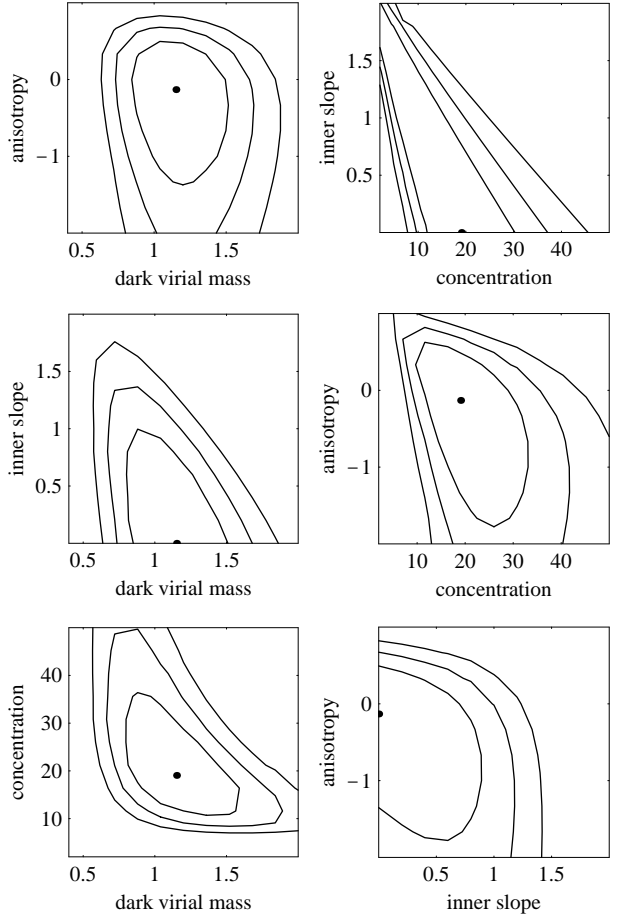}
\end{center}
\caption{Cuts through the $1\,\sigma$, $2\,\sigma$ and $3\,\sigma$
probability contours in
the parameter space obtained from fitting $\sigma_{\rm los}$
and $(\log \kappa_{\rm los})^{1/10}$.
Circles (and half-circles) indicate the best-fitting parameters.
The mass is in units of $10^{15} M_{\sun}$.}
\label{allplots}
\end{figure}

As discussed in Appendix A, the sampling distributions of $\sigma_{\rm los}$
and $(\log \kappa_{\rm los})^{1/10}$ are independent, hence we can use the same $\chi^2$
minimisation scheme to find joint constraints following from
fitting both quantities. Using the total of $14$ data points at $R<80'$, we jointly
fit our four parameters, $M_v$, $\beta$, $\alpha$ and $c$. Contrary to the case
when only velocity dispersion was studied, the minimisation procedure now
converges. The minimum
is found at $M_v = 1.2 \times 10^{15} M_{\sun}$ (corresponding to dark matter
virial radius $r_v=92' = 2.7$ Mpc), $\beta=-0.13$, $\alpha=0$ and $c=19$
with $\chi^2/N = 6.1/10$. For a better visualisation of the constraints
obtained for our 4
parameters, we plot in Figure~\ref{allplots} the cuts through the 4-dimensional
confidence region in all six possible planes with probability contours corresponding
to $1\sigma$ ($68$\%), $2\sigma$ ($95$\%) and $3\sigma$ ($99.7$\%) i.e.
$\Delta \chi^2 = \chi^2 - \chi^2_{\rm min} = 4.72, 9.70, 16.3$, where
$\chi^2_{\rm min}=6.1$.

\begin{figure}
\begin{center}
    \leavevmode
    \epsfxsize=8cm
    \epsfbox[45 40 330 550]{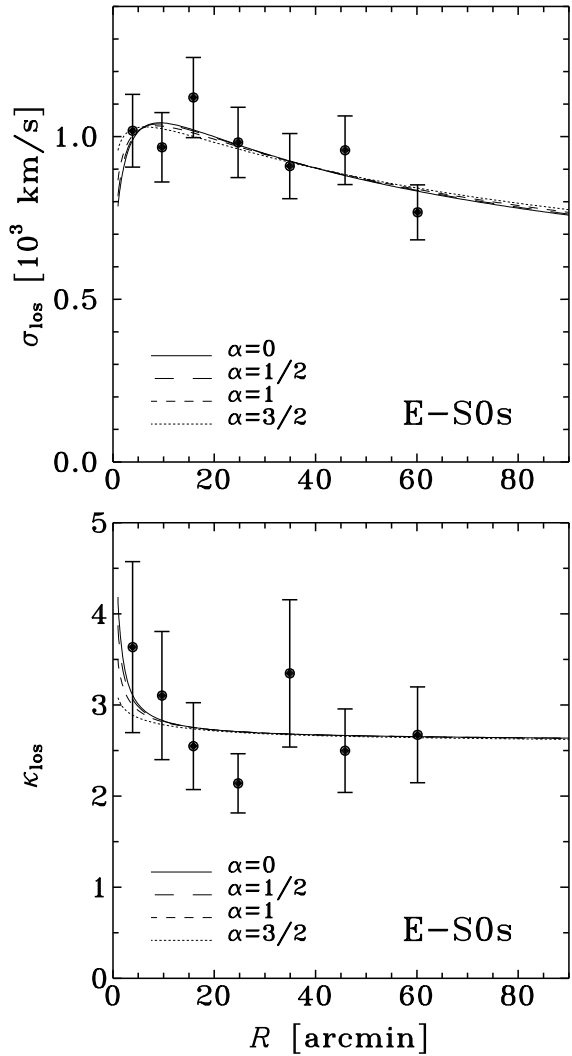}
\end{center}
\caption{The best-fitting line-of-sight velocity dispersion (upper panel) and
kurtosis (lower panel) profiles of E-S0 galaxies. The parameters of the models are
listed in Table~\ref{parameters}. The data are the same as in
Figure~\ref{vdkurte2}, except that only the 7 inner data points are shown in each panel.}
\label{dkfit}
\end{figure}

\begin{table}
\caption{Best-fitting parameters from joint analysis of $\sigma_{\rm los}$
and $\kappa_{\rm los}$ of E-S0 galaxies. $M_v$ is in units of $10^{15} M_{\sun}$.}
\label{parameters}
\begin{center}
\begin{tabular}{ccccc}
 $\alpha$ & $M_v$ & $\beta$ & $c$ & $\chi^2$ \\
\hline
 0   &  1.2 & $-0.13$ & 19 & 6.1\\
 1/2 &  1.2 & $-0.14$ & 14 & 6.2\\
 1   &  1.2 & $-0.16$ & 9.4 & 6.4\\
 3/2 &  1.2 & $-0.21$ & 4.9 & 6.9\\
\hline
\end{tabular}
\end{center}
\end{table}

\begin{figure}
\begin{center}
    \leavevmode
    \epsfxsize=8cm
    \epsfbox[45 40 330 550]{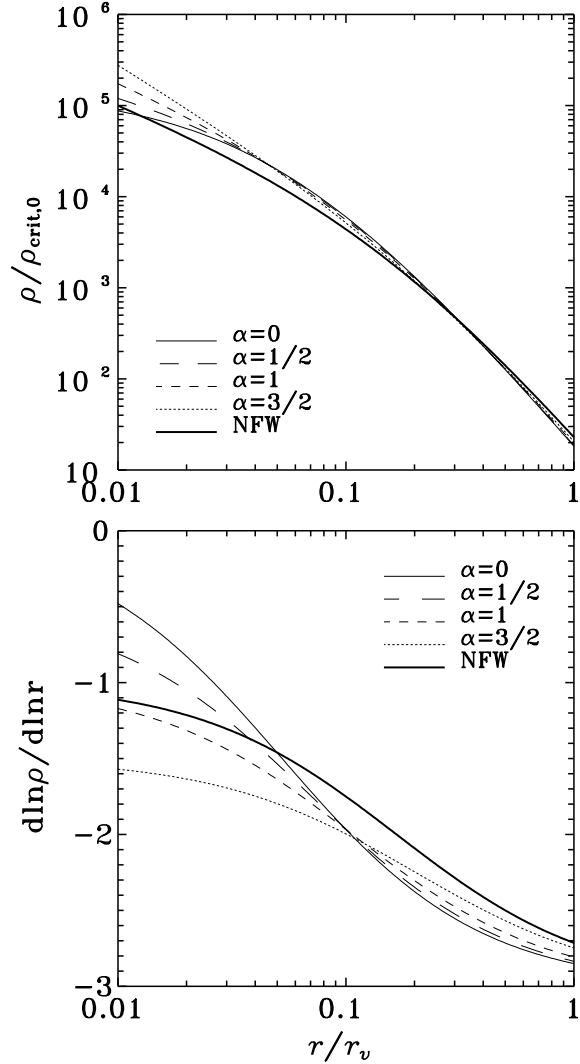}
\end{center}
\caption{The best fitting dark matter profiles with different inner slopes (upper panel)
and their effective slopes (lower panel) as a function of radial distance (in units of the
virial radius). The thick solid line in each panel shows the same quantities for the
standard NFW profile with $c=6$.}
\label{profslop}
\end{figure}

Although the cuts do not tell everything about the confidence region,
Figure~\ref{allplots} can be used to draw a number of interesting conclusions.
The most striking is the behaviour of the confidence region in the $c-\alpha$
plane shown in the upper right corner. Its shape shows that there is a strong
degeneracy between the two parameters and indeed almost equally good fits can be
obtained for the values of the inner slope other than $\alpha=0$. The best-fitting
values of the remaining
parameters (together with the corresponding $\chi^2$ value) when different
$\alpha$ are assumed are listed in Table~\ref{parameters}. The results confirm
that indeed $\alpha=0$ gives the best fit, but other inner slopes cannot
be excluded. The steeper the inner slope (the higher the value of $\alpha$),
however, the lower is the concentration required to provide good fits to the moments.

The velocity moments obtained with the sets of parameters listed in Table~\ref{parameters}
are shown in Figure~\ref{dkfit} together with the data; they overlap almost exactly.
The dark matter profiles following from equation (\ref{m6}) with the parameters from
Table~\ref{parameters} are plotted in the upper panel of Figure~\ref{profslop}. In the
lower panel we also show the logarithmic slopes of the profiles. In both panels our
best-fitting profiles are compared to the ``standard'' NFW profile with concentration
$c=6$ (as suggested by formula (\ref{c6a}) for the mass of the order of
$10^{15} M_{\sun}$). As can be seen in the upper panel, all our profiles overlap in
a wide range of radial distances $0.03\, r_v < r < r_v$, and are somewhat
steeper beyond $0.05\,r_v$ (see the
lower panel) than the standard NFW model deduced from cosmological $N$-body
simulations. Indeed, our best fitting NFW profile has
concentration $c=9.4$, higher by 50\% than the standard NFW profile for which
$c=6$ according to equation (\ref{c6a}).

\begin{figure}
\begin{center}
    \leavevmode
    \epsfxsize=8cm
    \epsfbox[45 40 330 550]{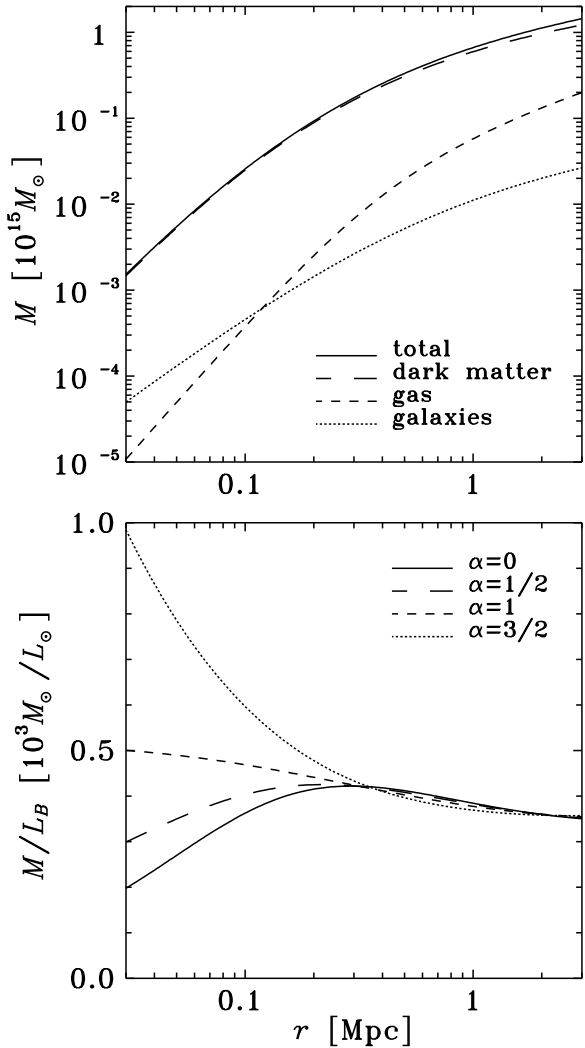}
\end{center}
\caption{Upper panel: The mass distributions for the best-fitting model ($\alpha=0$).
Lower panel: The mass-to-light ratios for the best-fitting models with different $\alpha$.
The dark matter virial radii are $r_v=2.7$ Mpc for all models shown.}
\label{massm2l}
\end{figure}

The remaining parameters $M_v$ and $\beta$ are better constrained. The three panels on
the left in Figure~\ref{allplots} show the probability contours for $M_v$ in the planes
with the other 3 parameters. We find the best estimate for the dark mass to be
$M_v = (1.2 \pm 0.4) \times 10^{15} M_{\sun}$ ($1\sigma$ error-bars). The anisotropy
parameter $\beta$ is very close to isotropic, the best fitting value $\beta
= -0.13$ gives $\sigma_\theta = 1.06\,\sigma_r$, i.e. the best-fitting
orbits of early-type galaxies are very weakly tangential, although fully consistent with
isotropy, while radial orbits are excluded at the $2\,\sigma$ level.

The upper panel of Figure~\ref{massm2l} compares the mass distributions in stars,
gas and dark matter with the total mass distribution for our best fitting model
with $\alpha=0$. At the dark matter virial radius $r_v=2.7$ Mpc, the total mass is
$1.4 \times 10^{15} M_{\sun}$. The mass in galaxies is only 2\% of the total mass,
the mass in gas is 13\% of the total (the galaxies thus represent less than
one-seventh of the baryonic mass), and the dark matter contributes the remaining
85\% of the total mass.
Therefore, the virial radius for the total mass is $0.85^{-1/3} = 1.06$ times
that of the dark matter, i.e. $97'$ or $2.9$ Mpc.

The cumulative mass-to-light ratio, i.e. the ratio of the total mass distribution to the luminosity
distribution in galaxies, is
\begin{equation}    \label{m2lratio}
	M/L_B = \frac{M(r)}{L_{\rm G}(r)} \ .
\end{equation}
where $M(r)$ is given by equation (\ref{c7}) with (\ref{c2}), (\ref{c5}) and (\ref{d5}),
while $L_{\rm G}(r) = M_{\rm G}(r)/\Upsilon$.
In the lower panel of Figure~\ref{massm2l} we show $M/L_B$ for our best-fitting models
for different $\alpha$, with parameters from Table~\ref{parameters}. The models
differ towards the centre and only for $\alpha=1$ does the distribution tend to a
constant value in this limit, since we have used the NFW profile to fit the luminosity
distribution of galaxies in Section~2. At radial distances larger than $0.3$ Mpc,
the cumulative mass-to-light ratio decreases slowly to reach $M/L_B \approx 351 M_{\sun}/L_{\sun}$
at the total mass virial radius $2.9$ Mpc.

\section{Discussion}

We studied the velocity moments of early-type galaxies in the Coma cluster
and used them to constrain the distribution of dark matter and velocity
anisotropy.
Our analysis differs from previous analysis of optical data (e.g. The \&
White 1986; Merritt 1987; den Hartog \& Katgert 1996; Carlberg et al. 1997;
van der Marel et al. 2000; Biviano et al. 2003; Biviano \& Girardi 2003),
in that
\begin{enumerate}
\item we have, for a single cluster, a larger sample of galaxies, which,
given their early morphological type, should be in dynamical equilibrium in
the cluster potential;
\item we remove pairs from the
computation of the velocity moments;
\item we include kurtosis in the analysis;
\item we model dark matter distribution using a generalised formula inspired by the
results of cosmological $N$-body simulations;
\item we include hot gas in our analysis.
\end{enumerate}

In comparison to studies based upon stacking of many clusters, our analysis
of the Coma cluster benefits from not having to introduce errors in any
stacking procedure, and from a cleaner removal of interlopers.
On the other hand,
the analyses of stacked clusters have the advantage of averaging out
particular inhomogeneities of individual clusters such as Coma, expected in
hierarchical scenarios of structure formation.
Indeed, the Coma cluster is known to have irregular structure both in
projected space (Fitchett \& Webster 1987; Mellier et al. 1988; Briel et
al. 1992) and velocity space (Colless \& Dunn 1996; Biviano et al. 1996).
In particular, the cluster has two central cD galaxies,
NGC 4874 and NGC 4889, of which the first
one is the central galaxy of the main cluster and the second probably
belonged to a subcluster which has recently merged with the main cluster
(e.g. Colless \& Dunn 1996). There are other subgroups, such as the one
associated with NGC 4839 at around $40'$ from NGC 4874, close
enough to the cluster centre to contribute to our analysis.

The question whether the E-S0 sample is relaxed and how the existing
substructure may affect our results can only be fully addressed
by cosmological $N$-body simulations including galaxy formation, where all
3-dimensional information would be available.
Although such an analysis has not yet been performed,
the effect of the incomplete virialisation of structures of dark matter
particles seen in cosmological simulations on the estimates of the mass
of a single cluster through the Jeans equation has been addressed
by Tormen, Bouchet \& White (1997). They showed (see the bottom row of their
Fig. 17) that even for significantly perturbed
haloes the mass $M(r)$ at distances larger than 2\% of the virial radius
inferred by the proper Jeans analysis is within 30\% (r.m.s.) of the true
mass and departs from it by less than 20\% (r.m.s.)
for average or relaxed haloes.
Since the dark matter distribution is known to possess more
substructure than is observed in the galaxy distribution (cosmological simulations
predict many more Milky Way satellites than are observed, see e.g. Moore et al. 1999)
and since structures cease to grow sooner in flat universes with a cosmological
constant, in comparison with analogous structures growing within the
Einstein-de Sitter model (assumed in the simulations by Tormen et al.), they
are more regular today (see Thomas et al. 1998) than shown
in the first three columns of the bottom row of Figure~17 in Tormen et al.
We therefore believe that the discrepancy between our derived mass and the
true cluster mass due to substructure and departures from equilibrium is
significantly smaller than the uncertainty due to sampling errors of the
velocity moments.

Our results for the
dark and total mass of the cluster are consistent with previous estimates. Using
a combination of X-ray and optical data Hughes (1989) found for his preferred model
a total mass within $3.6\,h_{70}^{-1}\,\rm Mpc$ (where we used the notation
$H_0 = 70\,h_{70} \, \rm km \, s^{-1}\,Mpc^{-1}$) to be $(1.3 \pm 0.2) \times 10^{15}
\,h_{70}^{-1}\,M_{\sun}$, but
a much wider range of masses if more general mass distributions were allowed.
Briel et al. (1992) using {\sf ROSAT} observations of Coma and assuming hydrostatic
equilibrium of the gas derived a mass within the same radial distance to be
$(1.3 \pm 0.4) \times 10^{15} \,h_{70}^{-1}\,M_{\sun}$.
Also, by analysing the infall patterns around Coma, Geller, Diaferio \& Kurtz
(1999) find a mass of $(1.7\pm0.4) \times 10^{15}\,h_{70}^{-1}\, M_{\sun}$
within the same distance to the cluster centre as above.
Extrapolating our results to this
radial distance, we find an enclosed mass of
$(1.6 \pm 0.5) \times 10^{15}\,h_{70}^{-1}\, M_{\sun}$, in good
agreement with these three earlier estimates.

We find a strong degeneracy between the inner slope and the concentration of
the dark matter profile, with many combinations
of the two reproducing our velocity profiles almost equally well. In the range of
inner slopes $0 \le \alpha \le 3/2$ we find that the best-fitting models have
the two parameters related almost linearly as $c=19 - 9.6\, \alpha$ with very little
variation in the remaining fitting parameters, $M_v$ and $\beta$.
The particular shape of
the degeneracy between $c$ and $\alpha$ is due to the specific properties of the family of
dark matter profiles used in this analysis, coupled with our lack of velocity
data at radii smaller than the scale radius ($r_s$) of the dark matter.
Using smaller radial bins would allow us to probe smaller radial distances,
but at the expense of larger sampling errors in the velocity moments.
As is clear from Figure 7, the
best-fitting dark matter profiles obtained for different inner slopes are almost
the same for a wide range of distances (larger than 3\% of the virial radius)
and this can only be achieved with the combination of the parameters as
given in Table 1.
Note that our best fit to the data for the flat dark matter density profile
might mean that the data may be even better fitted with the unphysical model
where the
dark matter density profile rises with radius in the cluster centre,
i.e. $\alpha < 0$ in equation (\ref{m6}).

Our best-fitting NFW profile ($\alpha=1$) has a concentration $c=9.4$ (in agreement
with the above formula), 50\% higher than $c=6$ found in $N$-body simulations.
It should be kept in mind, however, that the concentration parameters in the simulations
are subject to substantial scatter and that the formula (\ref{c6a}) (giving $c=6$) at
masses of the order of $10^{15} M_{\sun}$ is actually an extrapolation of results obtained
for lower mass haloes (see Bullock et al. 2001). However, the ``standard'' NFW model with
$c=6$ is still within our $1\sigma$ confidence region in Figure~\ref{allplots}.

In comparison, fits of the NFW profile to X-ray data of clusters, assumed isothermal,
by Ettori \& Fabian (1999), rescaled by Wu \& Xue
(2000), as well as by Sato et al. (2000), both yield $c \simeq
4$ for the mass that we find for Coma within the virial radius (the latter
after correction from $\Delta_{\rm c} = 200$ to $\Delta_{\rm c} = 102$), which is only
within our $2\,\sigma$ confidence region.
It may be that the assumption of isothermal hot gas causes a lower
concentration parameter.

In an analysis similar in spirit to ours,
Biviano et al. (2003) report $c = 4\pm2$ for their stacked
ensemble of 59 {\sf ENACS} clusters, fitted to an NFW model for the total mass
density. Similarly, Biviano \& Girardi (2003) show that NFW models with $c = 5.5$
are consistent with a stacked ensemble of 43 {\sf 2dFGRS} clusters.
Moreover, the {\sf ENACS} and {\sf 2dFGRS} clusters analysed in both studies
are on average less massive (i.e. with lower velocity dispersions) than Coma,
and given that cosmological simulations find a decreasing $c$ for increasing
mass, the discrepancy with our result is even stronger.

The difference of these four studies
with our result may be explained by the fact that the authors mentioned above fit the
\emph{total} mass density profile to the NFW form,
while our fit was done  for the distribution
of the dark component only.
If the gas is distributed similarly in the X-ray, {\sf ENACS} and {\sf 2dFGRS}
clusters as
in Coma, i.e. it has a flat inner core, and the gas on average contributes
a substantial part of the total mass, fitting the total mass distribution may result
in flatter profiles than the dark haloes really have.
The discrepancy may be also caused by the exclusion of central cDs by Biviano
et al. (2003),
although we feel that removing one or two galaxies from our central bin of 39
galaxies will not affect our results.
Alternatively, their mass model (obtained with
the assumption of isotropic orbits) may not
be coherent with the kurtosis profile of their stacked cluster.

Our best fit models have orbits that are very close to isotropic, as is
expected, in the central regions, where the two-body relaxation time for the
galaxy system is considerably smaller than the age of the Universe, and also
because
cosmological simulations indicate that dark matter particles typically
have isotropic orbits in the centres of clusters (Thomas et al. 1998;
Huss, Jain \& Steinmetz
1999). In other words, we find no significant anisotropy bias for the
galaxies relative to the expectations of isotropy for the dark matter.
Note that if we force isotropic orbits for the galaxies, we then
obtain similar constraints for the
inner slope, concentration parameter and mass of the dark matter component
within the virial radius as those shown in Figure~\ref{allplots}.

One might wish to better reproduce the kurtosis profile. It is clear from
Figure~\ref{dkfit} that, while the velocity dispersion profile
is well reproduced,
there is room for improvement for the kurtosis profile.
However, as discussed in Section~4, the kurtosis
is mainly sensitive to the velocity anisotropy, which was modelled here by a single
constant parameter. Besides, the curves shown in Figure~\ref{dkfit} are the
results of joint fitting of both moments and do not aim at reproducing the kurtosis alone.
In spite of this, the inclusion of the  kurtosis in our analysis allowed us
to constrain the velocity anisotropy and other parameters of the model.

{}From our best estimate of the mass distribution, the baryons (galaxies and
gas) contribute 15\% of the total mass at the dark matter
virial radius $r_v=2.7$ Mpc (see Figure~\ref{massm2l}). If we assume that the
cluster content is representative
of the Universe as a whole, we can use this baryon fraction to estimate the
density parameter $\Omega_0$ (see e.g. White et al. 1993). Taking the baryonic
density parameter at its currently best value from nucleosynthesis $\Omega_b=0.04$
(with $h=0.7$) we obtain $\Omega_0=0.26 \pm 0.09$ where the error comes from
our 30\% uncertainty in the dark virial mass value, 20\% uncertainty in the
gas mass (as estimated by White et al. 1993) and 10\% error in the $\Omega_b$
value (Burles, Nollett \& Turner 2001).

Similarly, we may assume that clusters are good tracers, within their virial
radius, of the ratio of mass to blue luminosity.
Given that the closure mass-to-light ratio (critical density over luminosity
density) in the blue band is roughly $1100\,h_{70}$ to 10\% accuracy,
consistent with the
recent estimates of luminosity density from the {\sf 2dFGRS} (Norberg et
al. 2002) and {\sf SDSS} (Blanton et al. 2001) galaxy surveys (after
correction to the blue band), our mass-to-light ratio within the virial
radius of $351\,h_{70}$ with 30\% accuracy yields
$\Omega_0 = 0.32 \pm 0.1$.

Combining these two estimates of the density parameter, we arrive at
$\Omega_0 = 0.29 \pm 0.1$ in excellent agreement with other determinations, for
example the recent value obtained from the {\sf WMAP} CMB experiment (Spergel et
al. 2003).

\section*{Acknowledgements}

We wish to thank Y. Hoffman and the anonymous referee for useful comments and
suggestions. We acknowledge hospitality of the Institut d'Astrophysique
de Paris where EL{\L} benefited
from the Jumelage exchange programme as well as the NATO Advanced Fellowship and
where most of this work was done.
Partial support was obtained from the Polish State Committee for
Scientific Research within grant No. 2P03D02319.
This research has made use of the NASA/IPAC Extragalactic Database ({\sf NED}) which is
operated by the Jet Propulsion Laboratory, California Institute of Technology,
under contract with the National Aeronautics and Space Administration. We have also
used the {\sf LEDA} database ({\tt http://leda.univ-lyon1.fr}) and {\sf SIMBAD}
operated at CDS, Strasbourg, France, as well as
NASA's Astrophysics Data System Service ({\sf
ADS}).

\appendix
\section{The sampling distributions}

The validity of modelling presented in this paper rests on proper estimation of
the velocity moments and their errors from observations. If their sampling
distributions tend to normality, the statistics obtained from a sample (e.g.
moments) can be characterised by an expectation value and a `standard' error
(Stuart \& Ord 1994). For simplest statistics, like the second central
moment, those can be calculated exactly from the population moments assuming
the population properties. Since we are interested here also
in quantities whose standard errors are not easily calculated analytically, we
resort to Monte Carlo methods.

The most natural estimators of the variance and kurtosis from a sample of
$n$ line-of-sight velocity measurements $v_i$ are
\begin{equation}    \label{app1}
	S^2 = \frac{1}{n} \sum_{i=1}^n (v_i - \overline{v})^2
\end{equation}
and
\begin{equation}    \label{app2}
	K = \frac{\frac{1}{n} \sum_{i=1}^n (v_i - \overline{v})^4}{(S^2)^2}
\end{equation}
where
\begin{equation}    \label{app3}
	\overline{v} = \frac{1}{n} \sum_{i=1}^n v_i
\end{equation}
is the mean of galaxy velocities in the sample.

To investigate the distribution of these estimators for our binning of galaxies,
i.e. when $n=39$, we ran Monte Carlo simulations by selecting ${\cal N}=10^4$ times
$n=39$ numbers from a Gaussian distribution with zero mean and dispersion of unity.
Since velocity distributions of gravitationally bound objects in general do not
dramatically depart from a Gaussian, this is a sufficient approximation for
constructing unbiased estimators of moments.

\begin{figure}
\begin{center}
    \leavevmode
    \epsfxsize=8.5cm
    \epsfbox[88 4 316 292]{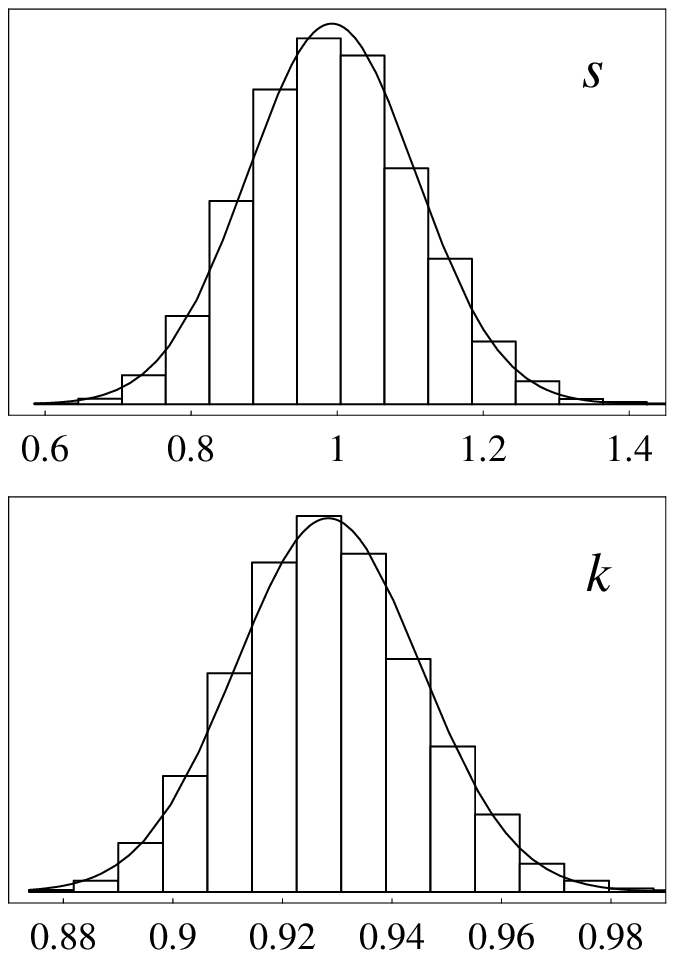}
\end{center}
\caption{Example of sampling distributions of $s$ and $k$
obtained from a Monte Carlo simulation of sampling from a Gaussian parent distribution
with $n=39$ and ${\cal N}=10^4$.}
\label{sampling}
\end{figure}

For each of the ${\cal N}$ samples we compute the statistic $\theta^\star_j$, namely $S^2$ or $K$
according to prescriptions given by equations (\ref{app1}) and (\ref{app2}). The Monte
Carlo estimate of our statistic is then the mean of all values obtained
\begin{equation}    \label{c8}
	\theta^\star = \frac{1}{{\cal N}} \sum_{j=1}^{\cal N} \theta^\star_j
\end{equation}
and its variance is
\begin{equation}    \label{c9}
	{\rm var} (\theta^\star) = \frac{1}{{\cal N}-1} \sum_{j=1}^{\cal N}
        (\theta^\star_j - \theta^\star)^2 .
\end{equation}
We find that the best estimates obtained in this way are biased, especially for kurtosis
(it is interesting to note, that the value of kurtosis is underestimated by a few percent
even for much larger samples with $n$ of the order of few hundred).
In addition, while the sampling distribution of velocity dispersion is Gaussian to
a very good approximation, the one for kurtosis is strongly skewed. Using this knowledge
we construct unbiased and Gaussian-distributed estimators of line-of-sight
velocity dispersion $s$ and kurtosis-like variable $k$
\begin{equation}    \label{app4}
	s = \left( \frac{n}{n-1} S^2 \right)^{1/2}
\end{equation}
\begin{equation}    \label{app5}
	k = \left[ \log \left( \frac{3}{2.75} K \right) \right]^{1/10} .
\end{equation}
The factor $n-1$ in equation (\ref{app4}) is the well known correction for bias when
estimating the sample variance, valid independently of the underlying distribution.
In (\ref{app5}) the factor $3/2.75$ corrects for the bias in the kurtosis estimate,
i.e. unbiased estimate of kurtosis is $K' = 3K/2.75$, while
the rather complicated function of $K'$ assures that the sampling distribution of $k$ is
approximately Gaussian. Examples of sampling distributions of $s$ and $k$ from our
Monte Carlo simulation are shown in Figure~\ref{sampling}. We find that the standard errors
in the case of $s$ are of the order of 11\% (in agreement with an analytic result
derivable with the assumption of Gaussian velocity distribution)
while in the case of $k$ are approximately 2\%.

The measured values of $\sigma_{\rm los}$ and $\kappa_{\rm los}$ calculated from our
velocity data using equations (\ref{app4})-(\ref{app5}) and (\ref{app1})-(\ref{app2})
are shown in Figure~\ref{vdkurte2}. The $1\sigma$ error bars for the velocity dispersion
are $0.11 s$. The values of kurtosis are $K' = 3K/2.75$ with approximate $1\sigma$ error
bars propagated from the 2\% error in $k$.

\begin{figure}
\begin{center}
    \leavevmode
    \epsfxsize=8cm
    \epsfbox[40 40 330 300]{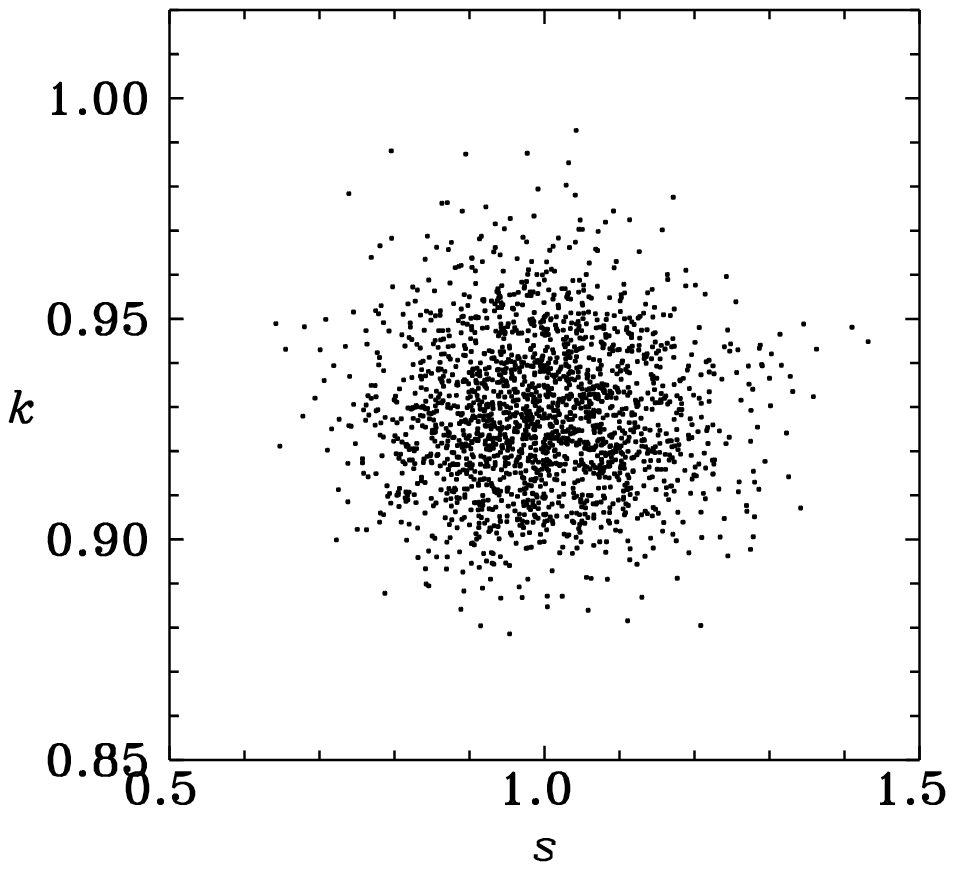}
\end{center}
\caption{The joint distribution of $s$ and $k$
obtained from a Monte Carlo simulation of sampling from a Gaussian parent distribution
with $n=39$ and ${\cal N}=10^4$. Only 2000 points are shown to reduce the size of the preprint.}
\label{sample}
\end{figure}

It is also important to check whether the sampling distributions of the two statistics
are independent. In general, the covariance between, e.g. even moments derived from the
same sample does not vanish, even for Gaussian distributions (Stuart \& Ord 1994).
However, the lowest order term is expected to decrease with the size of the sample $n$.
To check whether $n=39$ in our case is indeed large enough to assure independence of
$s$ and $k$, we construct the joint sampling distribution of the two statistics.
The joint distribution in the form of ${\cal N}$ points
with coordinates given by $(s,k)$ pairs calculated from each sample is presented in
Figure~\ref{sample}. As indicated by the Figure the variables are very weakly correlated,
we find the correlation coefficient of $|\varrho| \le 0.02$.

To check the behaviour of $s$ and $k$ for velocity distributions departing from gaussianity, we
repeated the Monte Carlo simulation, again sampling $n=39$ numbers from a Gaussian distribution,
but modifying each sample by removing the 6 most inner points and adding 3 uniformly
distributed in the range $(1\sigma, 2\sigma)$ and 3 in the range $(-2\sigma, -1\sigma)$.
Such distributions have unbiased kurtosis estimates of the order of $K'=2.2$, close to the
lowest value obtained from our data (and most strongly departing from the Gaussian value
of 3).
We find that estimators $s$ and $k$ are again Gaussian-distributed to a very good
approximation and very weakly correlated with $|\varrho| \le 0.07$. The sampling
distributions of $s$ and $k$
as well as the joint distribution look very similar to the purely Gaussian case.
The correlation coefficient can increase significantly only in the presence of additional
outliers (more numerous than predicted by the Gaussian distribution) in the range
$(2\sigma, 3\sigma)$ and $(-3\sigma, -2\sigma)$. We have checked that the number of galaxies
with velocities in this range is 1 or 2 in each bin, in excellent agreement with the Gaussian
prediction.

We can therefore assume that, to a good approximation, all our data points measuring
velocity dispersion and kurtosis
are independent, which justifies the use of standard $\chi^2$ minimisation to fit the models
to the data.


\begin{thebibliography}{}

\bibitem[]{ar} Arnaud M., Aghanim N., Gastaud R. et al. 2001, A\&A, 365, L67
\bibitem[]{bart} Bartelmann M., 1996, A\&A 313, 697
\bibitem[]{bm} Binney J., Mamon G. A., 1982, MNRAS, 200, 361
% \bibitem[]{bt} Binney J., Tremaine S., 1987, Galactic Dynamics. Princeton
%     Univ. Press, Princeton, chap. 4.
\bibitem[]{bi} Biviano A., 1998, in Mazure, A. et al., eds, Proc. Marseilles Meeting,
	Untangling Coma Berenices: A New Vision of an Old Cluster, Word Scientific,
    Singapore, p. 1, astro-ph/9711251
\bibitem[]{biv96} Biviano A., Durret F., Gerbal D., Le F\`evre O., Lobo C.,
Mazure A., Slezak E., 1996, A\&A, 311, 95
\bibitem[]{bktm} Biviano A., Katgert P., Thomas T., Mazure A., 2003, in Avila-Reese
   V. et al., eds, Proc. Cozumel Conference, Galaxy Evolution: Theory and Observations,
   RevMexAA, in press, astro-ph/0301343
\bibitem[]{bg} Biviano A., Girardi, M., 2003, ApJ, 585, 205
\bibitem[]{blanton01} Blanton M., Dalcanton, J., Eisenstein D. et al., 2001,
AJ 121, 2358
\bibitem[]{bhb} Briel U. G., Henry J. P., B\"{o}hringer H., 1992, A\&A,
    259, L31
\bibitem[]{bhfe}Broadhurst T., Huang X., Frye B., Ellis R., 2000, ApJ, 534, L15
\bibitem[]{bu} Bullock J. S., Kolatt T. S., Sigad Y., Somerville R. S.,
    Kravtsov A. V., Klypin A. A., Primack J. R., Dekel A., 2001,
    MNRAS, 321, 559
\bibitem[]{bur} Burkert A., 1995, ApJ, 447, L25
\bibitem[]{bnt} Burles S., Nollett K. M., Turner M. S., 2001, ApJ, 552, L1
\bibitem[]{carl} Carlberg R. G., Yee H. K. C., Ellingson E. et al., 1997,
ApJ, 485, L13
\bibitem[]{cd} Colless M., Dunn A. M., 1996, ApJ, 458, 435
\bibitem[]{dhk} den Hartog R., Katgert, P., 1996, MNRAS, 279, 349
\bibitem[]{ecf} Eke V. R., Cole S., Frenk C. S., 1996, MNRAS, 282, 263
\bibitem[]{ettori} Ettori S., Fabian A. C., 1999, MNRAS, 305, 834
\bibitem[]{fw} Fitchett M., Webster R., 1987, ApJ, 317, 653
\bibitem[]{fm} Fukushige T., Makino J., 1997, ApJ, 477, L9
\bibitem[]{gdk} Geller M. J., Diaferio A., Kurtz M. J., 1999, ApJ, 517, L23
\bibitem[]{hh} Henry J. P., Henriksen M. J., 1986, ApJ, 301, 689
\bibitem[]{huc} Huchra J. P., 1985, in Richter O. G., Binggeli B., eds,
    Proc. ESO Workshop on the Virgo Cluster of Galaxies. ESO, Munich, p. 181
\bibitem[]{hu} Hughes J. P., 1989, ApJ, 337, 21
\bibitem[]{hjs} Huss A., Jain B., Steinmetz M., 1999, MNRAS, 308, 1011
\bibitem[]{jvo} Jimenez R., Verde L., Oh S. P., 2003, MNRAS, 339, 243
\bibitem[]{js} Jing Y. P., Suto Y., 2000, ApJ, 529, L69
\bibitem[]{kg} Kent S. M., Gunn J. E., 1982, 87, 945
\bibitem[]{lo} {\L}okas E. L., 2002, MNRAS, 333, 697
\bibitem{lha} {\L}okas E. L., Hoffman Y., 2001, in Spooner N. J. C., Kudryavtsev V.,
    eds, Proc. 3rd International Workshop, The Identification of Dark Matter.
    World Scientific, Singapore, p. 121
\bibitem[]{lm} {\L}okas E. L., Mamon G. A., 2001, MNRAS, 321, 155
%\bibitem[]{ml} Mamon G. A., {\L}okas E. L., 2003, to be submitted to MNRAS
%\bibitem[]{mb} Magorrian J., Ballantyne D., 2001, MNRAS, 322, 702
\bibitem[]{mbl} McGaugh S. S., de Blok W. J. G., 1998, ApJ, 499, 41
\bibitem[]{mcl} McLaughlin D. E., 1999, ApJ, 512, L9
\bibitem[]{mell} Mellier Y., Mathez G., Mazure A., Chauvineau B., Proust D.,
1988, A\&A, 199, 67
\bibitem[]{mk} Merrifield M. R., Kent S. M., 1990, AJ, 99, 1548
\bibitem[]{mer2} Merritt D., 1987, ApJ, 313, 121
\bibitem[]{mgqsl} Moore B., Governato F., Quinn T., Stadel J., Lake G.,
    1998, ApJ, 499, L5
\bibitem[]{moore} Moore B., Ghigna S., Governato F., Lake G., Quinn T.,
    Stadel J., Tozzi P., 1999, ApJ, 524, L19
\bibitem[]{nfw} Navarro J. F., Frenk C. S., White S. D. M., 1997, ApJ,
    490, 493
\bibitem[]{norberg} Norberg P., Cole S., Baugh C. et al., 2002, MNRAS, 336, 907
\bibitem[]{sato} Sato, S. Akimoto, F., Furuzawa, A., Tawara, Y., Watanabe, M.,
 Kumai, Y., 2000, ApJ, 537, L73
\bibitem[]{sp} Spergel D. N., Verde L., Peiris H. V. et al., 2003, ApJ,
submitted, astro-ph/0302209
\bibitem[]{sr} Struble M. F., Rood H. J., 1999, ApJS, 125, 35
\bibitem[]{st} Stuart A., Ord K., 1994, Kendall's Advanced Theory of
    Statistics, Vol. I: Distribution Theory. Oxford Univ. Press, New York,
    chap. 10
\bibitem[]{tam} Tamura T., Makishima K., Fukazawa Y., Ikebe Y., Xu H., 2000,
ApJ 535, 602
\bibitem[]{tw} The L. S., White S. D. M., 1986, AJ, 92, 1248
\bibitem[]{thomas} Thomas P. A., Colberg J. M., Couchman H. M. P. et al.,
1998, MNRAS, 296, 1061
\bibitem[]{tbw} Tormen G., Bouchet F. R., White S. D. M., 1997, MNRAS, 286, 865
\bibitem[]{ts} Tully R. B., Shaya E. J., 1984, ApJ, 281, 31
\bibitem[]{tkd} Tyson J. A., Kochanski G. P., dell'Antonio I. P., 1998, ApJL, 498,
	107
\bibitem[]{vdm} van der Marel R. P., Magorrian J., Carlberg R. G., Yee H.
    K. C., Ellingson E., 2000, AJ, 119, 2038
\bibitem[]{so} White S. D. M., Navarro J. F., Evrard A. E., Frenk C. S., 1993,
    Nature, 366, 429
\bibitem[]{wnb} Williams L. L. R., Navarro J. F., Bartelmann M., 1999, ApJ,
	527, 535
\bibitem[]{wuxue} Wu X.-P., Xue Y.-J., 2000, ApJ, 529, L5



\end{thebibliography}
\end{document}